# An Exploration of Parallel Imaging System for Very-low Field (50mT) MRI Scanner

Lei Yang[1], Wei He[2], Sheng Shen[3], Yucheng He[4, 5], Jiamin Wu[4, 6], and Zheng Xu[2]

[1]National Center for Magnetic Resonance in Wuhan, State Key Laboratory of Magnetic Resonance and Atomic and Molecular Physics, Wuhan Institute of Physics and Mathematics, Innovation Academy for Precision Measurement Science and Technology, Chinese Academy of Sciences, Wuhan 430071, China.
[2]The school of Electrical Engineering, Chongqing University, Chongqing, 400044, China.
[3]A.A. Martinos Center for Biomedical Imaging, Department of Radiology, Massachusetts General Hospital, Charlestown, MA 02129 USA, and Harvard Medical School, Boston, MA, 02115, USA.
[4]Shenzhen Academy of Aerospace Technology, Shenzhen, China
[5]The school of Life Science, Northwestern Polytechnical University, Xian, China.
[6]Harbin Institute of Technology, Harbin, China.

**Correspondence to:**
   Zheng Xu, Ph.D., Prof.
   The school of Electrical Engineering
   Chongqing University, Chongqing, China
   Email: xuzheng@cqu.edu.cn

# INTRODUCTION

In recent years, with the development of high-performance hardware systems, advanced imaging sequences [1], and efficient image reconstruction and postprocessing methods [2], [3], the image quality of very-low field (VLF) magnetic resonance imaging (MRI) scanners has been significantly enhanced. In both the academic and industrial sectors, VLF MRI scanners have garnered a lot of research interest due to their ability to strike a balance between cost and imaging quality. In 2015, Sarracanie et al from Harvard University reported a 6.5mT VLF MRI scanner and conducted rapid brain imaging [4]. In 2019, our team introduced a 50mT VLF MRI scanner, and utilized in stroke diagnosis [5]. In 2020, O'Reilly et al from Leiden University medical center conducted 3D brain and extremity imaging on a 50mT VLF MRI scanner based on Halbach magnet [6]. In 2021 and 2022, Cooley from Harvard medical school [7] and our team [8] reported VLF MRI scanner which can provide point-of-care. In the industrial sector, Hyperfine, in collaboration with Yale University, first introduced a 64mT portable VLF MRI scanner in 2020, and it is designed for bedside monitoring and use in Neuro ICU [9].

However, owing to the diminished magnetic resonance signal resulting from the significantly lower main magnetic field intensity, VLF MRI scanners often necessitate averaging multiple-times scanning results to attain satisfactory image quality. This requirement prolongs the imaging duration, thereby compromising both patient comfort and the operational efficiency of the scanners. Parallel imaging (PI) techniques utilize the diverse sensitivity regions of phased-array coil units, coupled with specialized PI reconstruction algorithms, to achieve accelerated imaging [10]. While PI has established to be a mature method in high-field MRI scanner, its implementation in VLF MRI scanner necessitates a targeted redesign for both the phased-array coil and parallel imaging method.

Phased-array coil comprises numbers surface coil units and stands as the state-of-art in RF coil design [11]. Theses coils found widespread application in high-field scanners and play pivotal roles in various clinical scenarios, such as brain imaging [12], breast imaging [13], cardiac imaging [14], and lung imaging [15], [16]. Arrays used in these clinical settings typical range from four to more than 32 independent channels, with some advanced systems reached up to 128 channels [17] in high-field scanners. For phased-array coils suitable for PI, it is crucial that they exhibit diverse sensitivity regions across coil units. Notably, high-field scanner features a cylindrical magnet structure with main magnetic field parallel to the patient axis. In contrast, the VLF MRI scanners typically adopt bipolar structures [5], [9] and the main field transverse to the patient axis. Consequently, designing a phased-array coil for VLF MRI scanners requires consideration of the main magnetic field direction, operation frequency, and the target acceleration direction. This stands in contrast to directly applying the configuration of conventional high-field RF coils to VLF MRI scanners. Meanwhile, most of the reported phased-array coils designed for low field [18] or very-low field [19] MRI scanners are not specifically designed for PI.

Parallel imaging methods based on k-space focus on estimating the missing phase-encoding lines in k-space, and subsequently reconstructing the image. Notable techniques in this domain include simultaneous acquisition of spatial harmonics (SMASH) [20], a self-calibration technique for SMASH imaging (AUTO-SMASH) [21], variable-density sampling for AUTO-SMASH (VD-AUTO-SMASH) [22], generalized auto-calibrating partially parallel acquisitions (GRAPPA) [23] and nonlinear-GRAPPA (NL-GRAPPA) [24]. In recent years, the application of deep learning methods to parallel imaging has gained prominence, with approaches such as RAKI [25], DeepSPIRiT [26]. The coefficients utilized by these methods to estimate the missing phase-encoding lines are derived from the sampled auto-calibration signal (ACS) lines. However, the presence of noise in the ACS lines can significantly compromise the accuracy of the coefficient calculation. While NL-GRAPPA has introduced a nonlinear method to address the model bias, it is tailored for the high-field scenario where the MR signal is substantially larger than the noise. In contrast, VLF MRI scanners, even when operate in an electromagnetic shielding room, contend with considerable Gaussian white noise and the Johnson noise generated by the body. Thus, the PI method should be redesigned for the VLF MRI scanner account for the presence of noise.

In this paper, we introduce a novel parallel imaging technique for VLF MRI scanner, comprising of a specifically designed PI algorithm and a newly designed 8-channel phased-array coil. On the algorithm front, the undersampled k-space data from each coil unit undergoes denoising via a linear-prediction based Kalman filter. Subsequently, the denoised k-space data is nonlinearly mapped from the original space onto high-dimensional feature space, utilizing a polynomial feature mapping defined nonlinear frame. The missing phase-encoding lines in the original space are estimated by the acquired phase-encoding lines in the feature space. The proposed method termed linear-prediction based Kalman filter NL-GRAPPA (KF-NL-GRAPPA). In RF coil design, we have refined the structure of the phased-array coil based on insights gained from our previous work [27], and decoupling it using geometric overlap and low input impedance (LII) preamplifier.

Our team previously attempted an 8-channel phased-array coil, but did not achieve acceleration in imaging [27]. This work is the first attempt to implement PI technique on VLF MRI scanner. We evaluate the performance of the proposed PI technique through experiments involving healthy volunteer head imaging. A comprehensive evaluation is conducted by comparing the performance of the proposed KF-NL-GRAPPA method with the widely utilized GRAPPA and NL-GRAPPA method. The images results demonstrate that our proposed KF-NL-GRAPPA yields superior reconstructed image quality compared to the other methods. It achieves image quality akin to fully sampled images but in half scan time. Notably, the signal-to-noise ratio (SNR) of reconstructed image using KF-NL-GRAPPA surpasses that of the full sampled images, attributable to the denoising process. In the discussion section, we outline possible research directions, aiming to gain greater reduction factor in the PI process.

## METHOD

Parallel imaging relies on PI method and phased-array coil. In VLF MRI scanners, the MR signal is considerably weaker than that of the high-field counterparts, and the presence of interference can compromise the performance of the PI method designed for high-field MRI. VLF MRI scanner typically employ bipolar permanent magnets to generate a static main magnetic field perpendicular to the axis of the subject, contrasting with high-field MRI scanners where the main magnetic field is parallel to the axis of the subject. The method of this paper consists of two main parts: (1) the introduction of a VLF MRI parallel imaging algorithm; (2) design of an 8-channels phased-array coil for VLF MRI.

In the first part, the acquired k-space signal was denoised through applying the linear-prediction based Kalman filter. Subsequently, for a more reliable reconstruction of accelerated image results, the NL-GRAPPA method was employed. In the second part, we proposed an 8-channels phased-array coil specifically designed for VLF MRI scanner after researched the configuration and decoupling of it.

*A. Parallel Imaging Algorithm for VLF MRI*
  1) **Review of GRAPPA**

The GRAPPA [23] adopts a technique known as variable-density sampling method to obtain undersampled k-space. The outer part of k-space is undersampled along phase-encoding direction by a specified outer reduction factor (ORF), while the central part of k-space is fully sampled at the Nyquist rate referred to as ACS data. The k-space data of one acquired phase-encoding line and the missing lines between two acquired lines is called a block, as demonstrated in Fig. 1.

In general, GRAPPA method utilizes ACS data for calculating the linear estimated weights. The missing phase-encoding data in each coil is reconstructed by using the linear weighted combination of acquired undersampled data from the neighborhood encompassing all coils. The dimensionality of the neighborhood can be one, two or three, but the most used is the two-dimensional neighborhood. The mathematical expression of GRAPPA with two-dimensional neighborhood is:

$$S_j(k_x, k_y + m\Delta k_y) = \sum_{l=1}^{N} \sum_{a=1}^{D_x} \sum_{b=1}^{D_y} n_{j,m}(l,a,b) S_l(k_x + a\Delta k_x, k_y + bR\Delta k_y) \quad (1)$$

where, $j$ denotes the number of target coil. $S_j(k_x, k_y + m\Delta k_y)$ are the missing k-space points to be estimated in the target coil. $m = 1, \ldots, R-1$ and $k_y + m\Delta k_y$ means the phase-encoding position of the missing line. *l* counts through the number of the coil units, and *N* denotes the total number of the coil units of a phased-array coil. $S_l(k_x + a\Delta k_x, k_y + bR\Delta k_y)$ refer to the acquired data points in the undersampled k-space of the *l*-th coil unit. $n_{j,m}(l,a,b)$ are the linear estimated weights. $R$ means the ORF. $a$ and $b$ transverse the acquired neighboring k-space data in $k_x$ and $k_y$ direction, and $k_x$ and $k_y$ represent the coordinates along the frequency- and phase-encoding directions in the k-spaces, respectively. $D_x$ and $D_y$ are the size of the neighbourhood along the frequency- and phase-encoding directions, respectively.

The mathematical expression (1) of the GRAPPA method can be represented by the neighborhood interpolation network shown in Fig. 1. The missing data point (i.e., the left-hand side of (1) or the red dot in the interpolation network in Fig. 1) is calculated using a weighted linear combination of the acquired data points in the neighborhood (the right-hand side of (1) or the filled black circles in the interpolation network in Fig. 1). The GRAPPA method consists of two main steps: calibration and reconstruction. During the calibration step, the neighborhood interpolation network, as illustrated in Fig. 1, is applied to the ACS region. Since the GRAPPA method uses a variable-density sampling approach, the ACS region in the center of k-space is fully sampled at the Nyquist rate. This ensures that both the 'missing' data points and the acquired data points are known. This allows the calculation of the linear estimation weights in (1). In the reconstruction step, the interpolation network, using the previously calculated linear estimation weights, traverses the undersampled outer region of k-space to calculate the missing data points. Through this process, the GRAPPA method recovers all missing data points caused by undersampling.

The analysis conducted by Yuchou Chang [24] on the model variable error of GRAPPA highlighted that noise present in the ACS data could introduce nonlinear bias. The linear model of GRAPPA, when affected by this noise, leads to inaccurate calculation of weights, consequently degrading the quality of the reconstructed image.

The analysis was specifically tailored for high-field MRI scanner. In our scenario, the MR signal is significantly weaker than that of high-field ones, making Gaussian white noise more comparable. Simply correcting the model bias might not be sufficient here. Additionally, the working frequency here is also dramatically reduced, and the Johnson noise generated by human body further compromises the MR signal. Therefore, we proposed the KF-NL-GRAPPA parallel imaging algorithm for VLF MRI scenarios, as shown in Fig. 2. The linear-prediction based Kalman filter is utilized to denoise the acquired undersampled k-space data, and the denoised undersampled k-space data is subsequently reconstructed by using NL-GRAPPA.

**2) Signal interference suppression**
Kalman filter is an optimal estimation algorithm for extracting accurate information of a dynamic system with the presence of uncertainty. The dynamic system here is the MR signal, and the uncertainty is the interference in the MR signal. The Kalman filter is a recursive estimator, meaning that the estimated value of the current state can be calculated based on the estimated value of the previous state and the measurement value of the current state.

The state-space model of Kalman filter including state transition equation and observation equation, as given follows.

$$s_k = F s_{k-1} + n_k, \quad (2)$$
$$o_k = H s_k + m_k, \quad (3)$$

where, $s_k$ is the system vector at time $k$, $F$ is the state transition mode, and $n$ is the process noise which is subject to zero-mean normal distribution with covariance $W$. Equation (2) denotes the true state a time $k$ is evolved from the state at $k-1$. $o_k$ is the observation vector at time $k$, $H$ is the observation model which maps the true state space into the observed space, and $m$ is the observation noise which also is subject to zero-mean normal distribution with covariance $V$.

In this paper, the covariance $W$ of process noise can be obtained by calculating the prediction error variance. The periphery data of k-space contains very few MR signal as validated in our previous work [8], so it can be used to calculate the covariance $V$ of observation noise. The observation model is assumed to be $H = [0, \ldots, 0, 1]$.

The main body of Kalman filter consist of two parts: prediction and updating. The prediction part includes the state estimate prediction and estimate error covariance matrix, which are:

$$\hat{s}_{k|k-1} = F\hat{s}_{k-1|k-1}, \quad (4)$$

$$P_{k|k-1} = FP_{k-1|k-1}F^T + H^TWH, \quad (5)$$

where, $\hat{s}_{k|k-1}$ means the estimate of $s$ at time $k$ given observations up to and including at time $k-1$. $P$ is the estimation error covariance matrix.

The updating part includes the optimal Kalman gain, updated state estimate prediction and updated estimate error covariance matrix, which are:

$$K_k = P_{k|k-1}H^T(HP_{k|k-1}H^T + V)^{-1}, \quad (6)$$

$$\hat{s}_{k|k} = \hat{s}_{k|k-1} + K_k(o_k - H\hat{s}_{k|k-1}), \quad (7)$$

$$P_{k|k} = (1 - K_kH)P_{k|k-1}. \quad (8)$$

The k-space used in this paper is three dimensional. Initially, we unfold it into one dimensional k-space signal based on sampling order. Then, the system model can be described by linear-prediction coefficients (LPCs), which predict the current state from p previous measured data, where p is the predictor order [28]. The Levinson-Durbin method [29] was used to solving Yule-Walker equations to get the LPCs { $a_m$, $m = 1, \ldots, p$ }. The LPCs could replace the transition mode $F$ in (2) by extended to a $p \times p$ matrix. Then, we have:

$$F = \begin{bmatrix} 0 & 1 & \cdots & 0 & 0 \\ \vdots & \vdots & \ddots & \vdots & \vdots \\ 0 & 0 & \cdots & 1 & 0 \\ 0 & 0 & \cdots & 0 & 1 \\ -a_p & -a_{p-1} & \cdots & -a_2 & -a_1 \end{bmatrix}. \quad (9)$$

In the Kalman process, the initialization states are the first $l$ data points in the reshaped one-dimensional k-space signal. The initialization of estimate error covariance matrix ( $P$ ) is the matrix with diagonalization operation of $V$. Then, the true state (the signal after Kalman filter) of p-point data could be obtained by executing the prediction and updating part of Kalman filter. By repeating the same operation on the entire one-dimensional k-space signal data, we finally obtain the denoised 3D k-space data after reforming the filtered 1D signal.

3) **Nonlinear GRAPPA**
The input for NL-GRAPPA is the denoised k-space data from previous subsection. In the NL-GRAPPA, the missing k-space points to be estimated are termed target points and the acquired undersampled k-space points are referred to as source points as shown in Fig 2. The linear model between the target points and source points in the GRAPPA should be generalized into a nonlinear framework.

However, selecting the appropriate nonlinear function can be challenging. Inspired by the kernel method [30] from deep learning, we can find a proper feature mapping $\phi(\cdot)$ that nonlinearly maps the source points $S_{soc}$ from the input space onto $\phi(S_{soc})$ the feature space. The dimension of the feature space is usually much higher than that of the input space. The nonlinear framework can be written as:

$$S_{tag} = \sum_{i=1}^{N} n_i \phi(S_{soci}) \quad (10)$$

where, $S_{soci}$ means the $i$-th source point and $n_i$ represents the corresponding transfer factor for the $i$-th source point, which can be calculated using the least square method. $S_{tag}$ represents the missing data points in k-space, also referred to as target points, which can be fitted by using the source points. $N$ is the total number of source points required to fit a target point.

Although (10) still shows that the target points are a linear combination of the source points, it mathematically describes the nonlinear relationship between them due to the nonlinear feature mapping $\phi(\bullet)$. The goal of nonlinearly combining the source points in the input space is achieved by performing linear operation in the feature space.

The most general used kernel functions are the polynomial kernel and the Gaussian kernel. However, the Gaussian kernel can lead to overfitting of calibration data and imprecise parameter calculation in the nonlinear framework [31]. In addition, the linear combination of the source points in the input space should be reserved considering the successful application of the GRAPPA method. Furthermore, the feature mapping of the kernel function should have an explicit expression. Considering that the polynomial kernel satisfies these desired properties, the feature mapping of an inhomogeneous polynomial kernel in the following form [32] is selected here.

$$\phi(\mathbf{X}) = [1, \sqrt{2}x_1, \ldots, \sqrt{2}x_n, x_1^2, \ldots, x_n^2, \sqrt{2}x_1x_2, \ldots, \sqrt{2}x_{n-1}x_n]^T \quad (11)$$

where, vector $\mathbf{X}$ is composed of data points $x_i$, $i = 1, \ldots, n$. The vector $\phi(\mathbf{X})$ includes constant term and linear terms ($\sqrt{2}x_1, \ldots, \sqrt{2}x_n$) in the input space as desired. The second-order terms include the square terms within each coil ( $x_1^2, \ldots, x_n^2$ ), and the product terms between nearest neighbors ($\sqrt{2}x_1x_2, \ldots, \sqrt{2}x_ix_j, \ldots, \sqrt{2}x_{n-1}x_n$) in the k-space of each coil unit. $x_i$ and $x_j$ are the nearest neighbors in the k-space of each coil unit along $k_x$ (FE) direction. The above terms are repeated by the order across all coil units.

$$\begin{aligned}
S_j(k_x, k_y + m\Delta k_y) = \\
n^{(0)} \times 1 \\
+ \sum_{l=1}^{N} \sum_{a=1}^{D_x} \sum_{b=1}^{D_y} n_{j,m}^{(1)}(l, a, b) S_l(k_x + a\Delta k_x, k_y + bR\Delta k_y) \\
+ \sum_{l=1}^{N} \sum_{a=1}^{D_x} \sum_{b=1}^{D_y} n_{j,m}^{(2,0)}(l, a, b) S_l^2(k_x + a\Delta k_x, k_y + bR\Delta k_y) \\
+ \sum_{l=1}^{N} \sum_{a=1}^{D_x} \sum_{b=1}^{D_y} n_{j,m}^{(2,1)}(l, a, b) S_l(k_x + a\Delta k_x, k_y + bR\Delta k_y) \\
\times S_l(k_x + (a+1)\Delta k_x, k_y + bR\Delta k_y)
\end{aligned} \quad (12)$$

The feature mapping $\phi(\bullet)$ nonlinearly maps the source points from the input space onto the feature space. The target points can be estimated by linear combination of source points in the feature space. The nonlinear framework (12) can be obtained by plugging the feature mapping (11) into (1), where the same notations are used as in (1). The superscript of $n$ in (12) indicates the order of the term. For example, $n^{(0)}$ represents the constant term, $n_{j,m}^{(1)}$ represents the first-order (linear) term, $n_{j,m}^{(2,0)}$ means the square terms of the second-order, and $n_{j,m}^{(2,1)}$ means the product terms between nearest neighbors of the second-order terms. The first term on the right side of (12) corresponds to the constant term in (11). The second term corresponds to the first-order (linear) term in (11), which is the expression of GRAPPA method (1). The third term corresponds to the square terms within the k-space of each coil unit. The last term corresponds to the product of the nearest neighbors along the FE direction of each coil unit. These terms are then repeated in sequence across all $N$ coils.

Similar to (1), the mathematical expression in (12) can also be represented using a neighborhood interpolation network. In this network, the acquired data points (depicted as filled black circles in Fig. 1) are the source points in the feature space, while the target point to be estimated (shown as a red circle in Fig. 1) represents the missing data point in the original space. By applying the nonlinear framework described in (12), the target point can be calculated through a weighted combination of the feature-mapped acquired data points (source points).

The NL-GRAPPA method, like the standard GRAPPA approach, consists of two main steps: calibration and reconstruction. During the calibration step, the variable-density sampling ensures that all target and source points are fully sampled in the ACS region. The neighborhood is scanned along the PE and FE directions to cover all data points in the ACS region, constructing the target points matrix and source points matrix as shown in Fig. 2. The weight parameters of the nonlinear framework are then computed using a least-squares approach for complex numbers [33].

In the reconstruction step, the undersampled data in the outer k-space is recovered using the nonlinear framework with the calculated weight parameters and the acquired data points. We refer to the proposed method as KF-NL-GRAPPA.

*B. Design of An 8-Channels RF Phased-Array Coil*

After describing the proposed parallel imaging algorithm for VLF MRI, this subsection proceeds with the design of phased-array coil. It consists of two parts: (1) the configuration of an 8-channels RF phased-array head coil; (2) the decoupling of the phased-array coil. In the first part, we determined the geometrical structure of RF coil skeleton, the shape and position of each coil unit, and the number of turns of the coil unit. In the second part, the decoupling of RF coil was achieved by utilizing geometric overlap and LII preamplifier decoupling.

  1) **Configuration of phased-array coil**

The following experiment conducted aimed to inform the determination of the geometrical structure of the RF coil skeleton. Two solenoid coils with similar structures but different geometrical dimensions were utilized to acquire head scan images of the same healthy volunteer, as illustrated in Fig. 3. Both coil skeletons in Fig. 3(a) and (b) are elliptic cylinders, with the major and minor axes of the larger coil in Fig. 3(a) measuring approximately 255mm and 240mm, while those for the smaller coil in Fig. 3(b) are 225mm and 177mm. Adjacent to the images of the coils is the head scan images of the same healthy volunteer in Fig. 3. The regions of interest (ROIs) and the reference background region used for SNR calculation are shown in the green rectangle and red rectangle, respectively, in Fig. 3. The SNR was calculated by dividing the mean gray value in the ROIs by the mean gray value in the reference background region. The SNR of the image results acquired using the larger RF receiving coil is approximately 10.81, while the SNR using the smaller RF receiving coil is approximately 12.94. The experimental results indicate that a smaller coil leads to a higher filling factor and higher SNR in the scan image.

However, a coil with an excessively high filling factor may cause discomfort to the scanned subject. Furthermore, the impedance of the human body can significantly impact the matching of a coil with an excessive filling factor, leading to changes in tuning and matching (T&M) for different scanned subjects. Considering all these factors, the coil skeleton was designed to closely fit the head, striking a balance between SNR performance and comfort. The coil skeleton comprises a semi-ellipsoid and a semi-elliptical cylinder. The geometric dimensions of the semi-ellipsoid are 240mm, 200mm, and 50mm, while those of the semi-elliptical cylinder are 240mm, 200mm, and 160mm.

The static main magnetic field of an VLF MRI scanner is usually generated by a bipolar permanent magnet with a direction perpendicular to the axis of the subject. In our VLF MRI scanner, we use an H-shaped permanent magnet, as depicted in Fig. 4(a) and (b). The subject's axis aligns with the Y direction, and the main static magnetic field extends along the Z direction. To ensure that the radiofrequency magnetic field is orthogonal to the main static magnetic field, the RF magnetic field should be along the X or Y direction, as illustrated in Fig. 4(c).

The 8-channels phased-array RF coil proposed in our previous work [27] has a drawback of losing image information in the top and bottom region along Z direction, as detailed in Discussion Section.

Each coil unit in this paper is a surface coil, but two 8-shaped coil units are placed on the top and bottom of the RF skeleton along the Z direction. According to the characteristics of surface RF coils [34], each coil unit should have similar size to ensure comparable sensitivity region and SNR across the entire ROI. The number of turns of each coil unit was determined by analyzing its relationship with the quality factor

(Q factor), using the method in [35]. For each coil unit of the phased-array coil, we wound it with different numbers of turns and measured its resistance $R_e$ and inductance $L_e$. The Q factor of each coil unit with varying turns was then calculated using (13), where $\omega_e$ denotes the Larmor frequency. To ensure that each coil unit of the phased-array coil achieves similar SNR and imaging depth, the number of turns for each coil unit was empirically selected to be 4, based on the Q factors. The configuration of the 8-channel phased-array RF coil is illustrated in Fig. 5(a), while the number and relative positions of each coil unit are depicted in Fig. 5(b).

$$Q = \frac{\omega_e L_e}{R_e} \quad (13)$$

The sensitivity maps of CH1, CH3, CH5, and CH7 are illustrated in Fig. 6. The sensitivity regions in the top and bottom regions along Z direction are completed by introducing two 8-shaped coils (CH7 and CH8) as shown in Fig. 6(d).

**2) Decoupling the phased-array coil**
The coupling between the RF coil units induces noise amplification, leading to a reduction in the SNR of each coil unit. Moreover, crosstalk between coil units diminishes the independence of the sensitive region for each coil unit, thereby limiting the achievable reduction factor of the parallel imaging algorithm.

The coupling of two coils could be demonstrated by Fig. 7. Circuit I and circuit II represent two coupled coils. $R_1$ and $R_2$ are the resistances of coil I and coil II, respectively. $L_1$ and $L_2$ denote the inductances of the two coils. $C_1$ is the capacitance in series with coil I, while $C_{2a}$ and $C_{2b}$ are in series with coil II. $R_{preamp}$ represents the input impedance of the LII preamplifier, and $L_m$ is the inductance in series with it. Considering the coupling of coil II, the output voltage signal of coil I can be expressed as

$$V_{output} = V_{signal} + (R_1 + j(\omega L_1 - \frac{1}{\omega C_1}))I_1 + i\omega M_{12} I_2 \quad (14)$$

where, $V_{signal}$ represents the voltage signal induced by coil I, $M_{12}$ is the mutual inductance between two coils, $I_1$ and $I_2$ signify the current in the coil I and coil II respectively. In the MRI system, $V_{output}$ of the left side of (14) is the voltage signal measured at the output port of coil I. The first item on the right side of (14) (i.e., $V_{signal}$) represents the pure MR signal, the second item represents the thermal noise of coil I itself, the third item is the thermal noise introduced by coil II to coil I through coupling. The objective of decoupling is to minimize the third item.

Minimizing the third item on the right side of (14) can be accomplished by minimizing $M_{12}$ or $I_2$. Minimizing $M_{12}$ is known as the geometric decoupling, achieved by overlapping the adjacent coils. The suitable overlapping area has been previously summarized for coils with regular structures, such as rectangles and circles [36]. For coils with irregular structures, the overlapping area can be determined by minimizing the S21 curve through adjusting the positions of the coils [37].

The coupling between nonadjacent coils can be minimized by reducing $I_2$. In the right part of Fig 7, the LII preamplifier is utilized to reduce the crosstalk between two coils. The input impedance of LII preamplifier proposed in our previous work [27] is very low (about 1 Ω). At this time, $L_m$ and $C_{2b}$ can be tuned to resonance in parallel by selecting an appropriate value for $L_m$. When parallel resonance occurs in coil II, the current $I_2$ is minimized.

The 8-channel phased-array coil was completed following geometric decoupling, and the finalized design proposed in this paper is shown in Fig. 8. The coil was wound using 0.03 mm * 2150-strand Litz wire.

The overlapping area between adjacent units can be adjusted according to S21 curve to minimize the coupling. In Fig. 8(a), there were no prefabricated wire troughs on the skeleton, allowing for slight adjustments to the position of coil unit. In the T&M circuit exhibited in Fig. 8(b), The inductance ($L_m$) in the red rectangular was tuned to resonance in parallel with the capacitance ($C_{2b}$) in the yellow rectangular next to it at designated operational frequency (2.323MHz).

We measured the coupling values, known as S21 parameters, between each coil unit and the other coil units at the designated operational frequency. The measurement method involved placing the 8-channel phased-array coil inside the magnetic cavity, for the two coil units targeted for measurement, they were connected to the measurement circuit and their S21 curves were obtained using a network analyzer. The remaining six channels were connected to the subsequent units of the VLF MRI scanner through their respective LII preamplifiers. The results of the measurements are presented in Table I.

The statistical data in Table I denotes that the adjacent coils are well decoupled by selecting proper overlapping area. The coupling effect between nonadjacent coil units is mitigated with the use of the LII preamplifier.

## RESULTS

We conducted head scanning experiments on healthy volunteer to validate the proposed KF-NL-GRAPPA method and 8-channels phased-array coil on a 50mT MRI scanner. Additionally, we compared the proposed KF-NL-GRAPPA method with the commonly used 2D-GRAPPA and the NL-GRAPPA.

*A. Experiment Methods*

The scanner was positioned in an electromagnetic shielding room, as shown in Fig 9(a), and the spectrometer (MRsolution EVO, Guiford U.K.) featured eight receiver channels. The center frequency of the 50mT permanent magnet with an H-shaped structure was 2.323MHz. The proposed 8-channels phased-array coil was connected to mid-amplifier through LII preamplifier, as shown in Fig 9(b).

A 3D gradient echo (GE3d) sequence with flip angle of 60 degree, a repetition time (TR) of 39ms, and an echo time of 17ms was used in the head scanning. The field of view (FOV) was set to 256×256×100, and the matrix size of k-space is 176×176×10.

The k-space was initially fully sampled at the Nyquist rate, and the fully sampled 3D k-space was subsequently undersampled. The outer of k-space was undersampled by ORF of 3 along the PE direction, and the missing PE line were filled with zeros. The number of ACS lines was 42 in the center of k-space. The size of neighbourhood used in this paper was two blocks along the PE direction and seventeen data points along the FE direction.

The subsequently undersampled k-space of each coil unit was restored applying the 2D-GRAPPA, NL-GRAPPA and the KF-NL-GRAPPA method. The reconstructed image and the ground truth image of each coil unit obtained from the restored and originally fully sampled k-space of each coil unit. The synthesized image of all coil units was acquired by using a sum-of-square method [36] on each coil image result.

*B. In-vivo Head Scanning Results*

In the experiment, the slice selection, PE, and FE gradient were along Y direction, X direction and Z direction, respectively. The image results of fully sampled, undersampled, GRAPPA restored, NL-GRAPPA restored and KT-NL-GRAPPA restored k-spaces are exhibited in Fig. 9(a), where "CH1" to "CH8" represent the image of each coil unit, and "Syn" denotes synthesized images.

The synthesized images in Fig. 10(a) indicates that the issue of missing image information in the top and bottom region along Z direction, as observed in our previous work [27], was resolved by incorporating two 8-shaped coils (CH7 and CH8).

The enlarged synthesized images are displayed in Fig 10(b), with the original image being the result of fully sampled k-space, referred to as the original image. The undersampled image is constructed from undersampled k-space, where missing PE lines are filled with zeros. The undersampled image exhibits wrap-around aliasing and low resolution, with a zoomed-in view of the aliasing presented in the blue rectangle next to it.

The lower three images in Fig. 10(b) are the synthesized images of restored k-spaces obtained using

GRAPPA, NL-GRAPPA, and the proposed KF-NL-GRAPPA. The detailed image results within the blue rectangle in the background are provided alongside. No visible aliasing is observed in the background region for all three reconstruction methods; however, the background noise is lowest in the KF-NL-GRAPPA restored image. Additionally, reconstruction artifacts are highlighted in the zoomed-in images of the signal region within the yellow rectangle. Based on these zoomed-in images, it can be qualitatively analyzed that the NL-GRAPPA method demonstrates better reconstruction performance than the GRAPPA method. Among these three methods, the proposed KF-NL-GRAPPA method exhibits the best performance, as shown in the lower right of Fig. 10(b), with the signal region having the smallest reconstruction artifacts and displaying image quality similar to the original image. The quantitative performance analysis of the three reconstruction methods is provided below.

The quantitative metrics adopted here are mean squared error (MSE), peak signal-to-noise ratio (PSNR), and structural similarity index (SSIM). In pMRI, MSE quantifies the difference between the reconstructed image and the fully sampled image, reflecting the average pixel-level error. A smaller MSE indicates a higher quality reconstructed image, while significant reconstruction artifacts will lead to a substantial increase in MSE. PSNR measures the SNR of the reconstructed image, offering an intuitive assessment of image quality. Higher PSNR values signify better image quality, meaning the reconstructed image is closer to the fully sampled image. SSIM evaluates the similarity between two images by measuring luminance, contrast, and structure. It detects the similarity and distortion between the reconstructed and fully sampled images. A higher SSIM value indicates a smaller discrepancy between the reconstructed image and the fully sampled image, thus indicating better image quality. In high-field pMRI [38-39], these three metrics—MSE, PSNR, and SSIM—are used collectively to quantitatively analyze the reconstruction image quality and the performance of various methods. The metrics for the three methods are presented in Table II.

According to Table II, the proposed KF-NL-GRAPPA method exhibits the smallest MSE value and the highest PSNR and SSIM values. Collectively, these three metrics indicate that the KF-NL-GRAPPA method delivers the best reconstructed image quality compared to the commonly used GRAPPA and NL-GRAPPA methods.

It is noteworthy that, unlike high-field MRI scanners [11], the SNR of the reconstructed image using the proposed parallel imaging technique is improved in the VLF MRI scanner. ROIs used for SNR evaluation are outlined by red rectangles in the different image results shown in Fig. 10(b). The reference background regions are indicated by green rectangles. The SNR was calculated by dividing the mean gray value in the ROIs by the mean gray value in the reference background region. The SNR in different regions and the time consumption for each image result are presented in Table III.

SNR1, SNR2, and SNR3 in Table III represent the SNR values for different image results in the three ROIs indicated in the Fig. 10(b). The proposed KF-NL-GRAPPA method achieves the highest SNR values across all three regions while consuming the least amount of time, as indicated in Table III.

The qualitative scan results and quantitative metrics for the healthy volunteer demonstrate that the proposed KF-NL-GRAPPA method, in conjunction with the 8-channel phased-array coil, ensures similar image quality with improved SNR as the fully sampled image, while requiring only half the scan time. The proposed KF-NL-GRAPPA method outperforms the commonly used GRAPPA and NL-GRAPPA methods.

## DISCUSSION

In this paper, we introduced the application of parallel imaging technology to VLF MRI scanners for the first time. We presented a parallel imaging method tailored for VLF MRI, termed as the KF-NL-GRAPPA algorithm. Additionally, we designed an 8-channel phased-array coil, and these two components together constitute the parallel imaging technique for VLF MRI scanners. We conducted tests on the proposed parallel imaging technique and compared our method with commonly used techniques in in-vivo human brain imaging. The results concluded that the proposed algorithm provides the best image quality among them. The reconstructed image exhibits similar quality with improved SNR compared to fully sampled images, all while utilizing only half the scan time.

*A. Key Factors for Enhancing SNR in Reconstructed Images*

As shown in the Table III, the SNR of the reconstructed image using the proposed parallel imaging technique is enhanced in the VLF MRI scanner, in contrast to high-field MRI scanners [11]. We believe that the improvement in SNR of VLF MRI scanner is due to the following three factors. Firstly, Kalman filter noise suppression: The Kalman Filter used in our method can effectively suppress Gaussian white noise in the electromagnetic shielding room and Johnson noise generated by the human body. By reducing noise, the features of the magnetic resonance signal are highlighted, leading to more accurate reconstructed images. Secondly, minimal noise from reconstruction: The parallel imaging method reconstructs only the missing MR signal, introducing minimal additional noise. The image results of VLF MRI scanner, even when fully sampled, have a much lower SNR compared to high-field MRI images. This means that the background noise intensity in VLF MRI is significantly higher than in high-field MRI. Consequently, the noise enhancement produced by the reconstruction algorithm is almost negligible compared to the background noise. Thirdly, relatively low reduction factor: The total reduction factor used in this study is 2.03, which is a relatively low value. This lower reduction factor helps maintain less noise amplification during image reconstruction.

*B. Validation of Adaptability of the Proposed Method*

We applied the proposed method to the phased-array coil designed in our previous work [27] to verify its adaptability. The image results of another healthy volunteer brain scan are exhibited in Fig 11.

The same quantitative performance metrics were included, and the proposed method was compared with GRAPPA and NL-GRAPPA. The MSE for GRAPPA, NL-GRAPPA, and the proposed KF-NL-GRAPPA were 18.09, 16.82, and 12.73, respectively. The PSNR values for these methods were 35.55 dB, 36.27 dB, and 37.39 dB, while SSIM scores were 0.90, 0.91, and 0.93. The images reconstructed using the proposed method exhibited the smallest MSE and the highest PSNR and SSIM among the three methods. The quantitative and qualitative results in Fig. 11 reaffirm this conclusion, demonstrating that the proposed method provides the most comparable image quality to the fully sampled images. However, some brain information is missing in the top and bottom regions along the Z direction, as noted in the previous section. Comparing Fig. 10 with Fig. 11, the 8-shaped coil units (CH7 and CH8) in the proposed 8-channel phased-array coil effectively recover the missing brain information, ensuring that the VLF MRI scanner can utilize the pMRI technique.

*C. Method to Improve the Reduction factor*

We believe that the reduction factor used in this paper does not reach the limit of the proposed method. It is limited by the following conditions which can be improved to reach higher reduction factors. These conditions are (1) linearity of gradient coil, (2) decoupling performance between coil units, (3) the optimization of the phased-array coil. The condition 3 will be discussed in detail in next subsection.

The essence of GRAPPA class method lies in utilizing combination of signals from multiple surface coils with different spatial sensitivities to generate spatial harmonics which should be generated by gradient encoding. In k-space, the combinations generate multiple data set with distinct offsets [20]. If the linearity of gradient coil is poor, the reconstruction error (reconstruction artifacts) may occur. In this case, the reduction factor of parallel imaging method needs to be turned down to ensure the reconstruction image quality.

Crosstalk between coil units in a phased-array coil can make the sensitivity regions of different coil unit to overlap, thereby reducing the spatial sensitivity variability. The variability of spatial sensitivity among different coil unit along the PE direction determines the maximum theoretical reduction factor. This underscores the importance of decoupling between coil units in parallel imaging. The decoupling methods used in this paper are geometric overlap and LII preamplifier decoupling. The geometric overlap can effectively cutoff the coupling between the adjacent coil units. Despite the LII preamplifier could eliminate the frequency splitting caused by nonadjacent coil units coupling, its decoupling performance may not meet expectations.

During the T&M process of coil units, it was observed that short-circuiting the output port of the T&M circuit of one coil unit (e.g., CH1) almost eliminated crosstalk in another coil unit (e.g., CH2). However, when connecting the output port of the T&M circuit of CH1 to the LII preamplifier, the coupling in CH2 was weakened but far from eliminated. Our previous work [27] also has also indicated that coupling increases with the input impedance of the LII preamplifier. The input impedance of our LII preamplifier is very low, measuring approximately 800 $m\Omega$, comparing with the LII preamplifier used in high-field MRI scanner whose input impedance is about 5 $\Omega$. However, the loaded RF coil in VLF MRI scanner has a significantly low resistance than the input impedance of LII preamplifier. The LII preamplifier must balance the performance of input impedance, gain, and noise figure, we cannot only pursue low input impedance and ignore other factors. In summary, the LII preamplifier might not be an ideal method to eliminate the crosstalk between nonadjacent coil units.

Applying novel decoupling technology to solve the coupling problem in VLF MRI is a valuable direction to investigate. Magnetic wall [40] approach combined with common-mode differential-mode (CMDM) resonators. The main strategy involves using two orthogonal loops with tuning capacitors to suppress electromagnetic coupling between multiple current modes in a CMDM quadrature array. This double-cross magnetic wall decoupling technique effectively reduces the mutual coupling between the coil units, enhancing SNR and overall imaging performance. The approach allows for better control of the decoupling process, providing a robust solution for designing phased-array coil. The clip-path conductors (CPCs) approach [41] is a novel geometric decoupling method for phased-array coils that has attracted significant interest. This technique involves placing CPCs in specific positions within the loop coil to reduce mutual inductance coupling between adjacent elements. By positioning the CPCs at the top-right and bottom-left of the coil loop, this method enables effective control over the decoupling performance through adjustments to parameters such as the conductor path length and segmentation capacitance.

*D. Method to Further Optimize the Phased-Array Coil*

The synthesized sensitivity region of all coil units must cover entire ROIs, and the spectrometer used in this paper has only eight receiver channels, so the flexibility of coil unit placement is limited. Each coil unit should have similar geometric structure to ensure comparable sensitivity region size and SNR. To achieve better geometric overlap decoupling performance, CH5 and CH6 in the proposed phased-array coil are larger than other coil units. Although the two 8-shaped coil (CH7 and CH8) can supplement the image information, it may be beneficial to replace them with multiple similarly structured RF coils with radiofrequency magnetic field components orthogonal to the main static magnetic field. Optimizing the location and structure of RF coil units can be achieved by incorporating more receiver channels. Further optimization in the location and structure of coil units can be explored in the following directions.

The possible parallel imaging direction of a basic phased- array coil is exhibited in Fig. 12. The PE direction should be consistent with the possible parallel imaging direction. The theoretical reduction factor of parallel imaging method increases with the number of coil units in the PE direction. Introducing more receiver channels to the spectrometer allows for more coil units to be arranged in the target PE direction, aiming for a higher theoretical reduction factor. Therefore, the location of coil units can be arranged as many coil units as possible along the target PE direction, under the condition that the synthesized sensitivity map can cover the entire target area. If each of the coil unit is the same simple regular shape, such as rectangle and circle, the size of the overlapping area that eliminates mutual inductance of adjacent coil unit has been calculated in [36].

The optimization of coil unit structure is a balance of SNR, uniformity of synthesized radiofrequency magnetic field, and the coupling degree. It is essential to ensure that the sensitivity regions of coil units along the PE direction should be well separated, indicated by minimizing the so-called geometry- or g-factor [10]. We believe that the reduction factor of the proposed KF-NL-GRAPPA method can be improved by optimizing the multi-channels phased-array coil, so this is another interesting area worthy researching.

# CONCLUSION

To summarize, this work is the first attempt to implement parallel imaging for VLF MRI scanner. Our proposed parallel imaging technique comprises the KF-NL-GRAPPA VLF MRI parallel imaging algorithm and an 8-channels phased-array coil. Through evaluation using healthy volunteer brain imaging experiments, we compared its performance with widely used GRAPPA and NL-GRAPPA methods. The image results demonstrates that the KF-NL-GRAPPA method provides superior image quality to other two methods. It can achieve similar image quality to fully sampled images in half scan time. Furthermore, it enhances SNR, comparing the reconstructed images and fully sampled images. While acknowledging the potential for further reduction factor improvement, we believe improving the decoupling using novel technology and optimizing the multiple channels (more than eight) phased-array coil are promising research direction.

# ACKNOWLEDGEMENTS

This work was supported by the National Natural Science Founding of China under Grant 52077023, the Shenzhen Science and Technology Innovation Commission under Grant CJGJZD20200617102402006, the Chongqing Science and Health Joint Project under Grant 2023MSXM016 and the Shenzhen Science and Technology Innovation Commission under Grant KJZD20230923114110019.

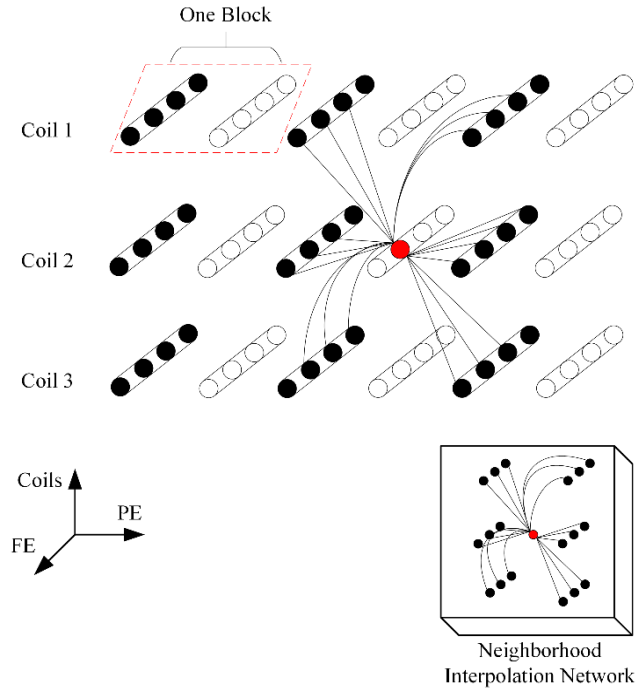

Fig. 1. Diagram of GRAPPA, where filled black circles represent acquired data points, hollow circles indicate missing data points, and the red circle marks the specific missing data point to be calculated. In this example, a phased-array coil with three coil units is used. The k-space is undersampled by a reduction factor of 2. The neighborhood, as illustrated in the interpolation network, consists of two blocks along the phase-encoding (PE) direction and three data points along the frequency-encoding (FE) direction. The missing data points in the unacquired PE line are estimated using a weighted linear combination of the acquired data points within the neighborhood, as depicted in the interpolation network.

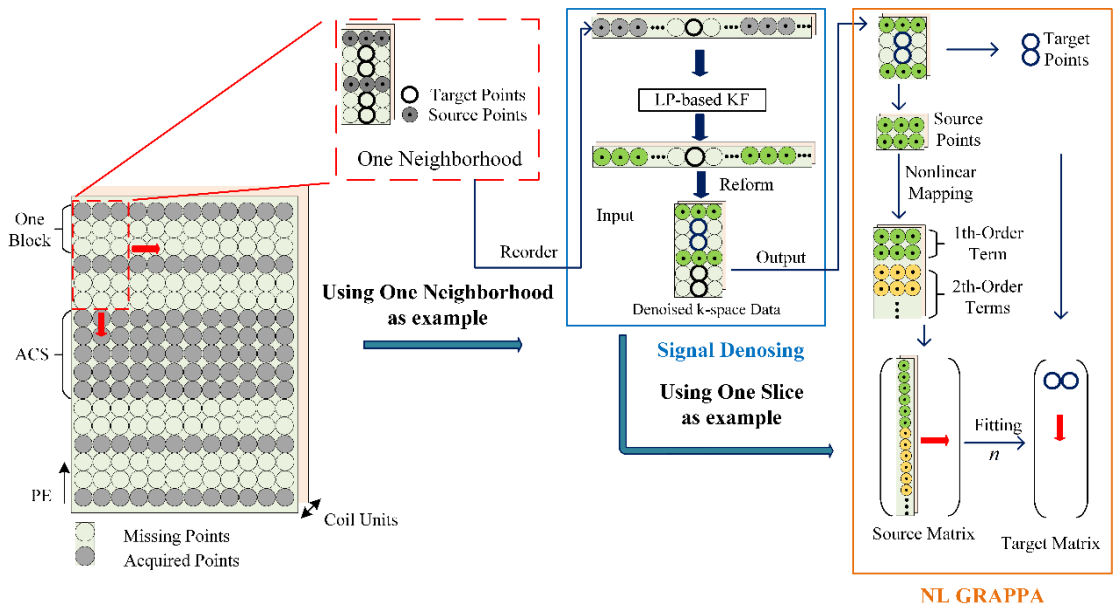

Fig. 2. Diagram of KF-NL-GRAPPA, where PE means phase-encoding and LP means linear prediction. The k-space is undersampled by ORF of 3. The neighbourhood, as shown in the left part with an enlarged view next to it, comprises two blocks along phase-encoding direction and three data points along frequency-encoding direction, serving as an example. The processing of KF-NL-GRAPPA is illustrated using a single neighbourhood. The 3D k-space is reordered into one-dimension, denoised using a LP-based KF, and subsequently reformed into 3D by restoring the original sampling order. The denoised k-space is inputted into the frame of NL-GRAPPA. The source points undergo a nonlinear mapping, where they are projected onto a feature space through a polynomial feature mapping. Estimations of the target

points in the original space are derived from the source points in the feature space. By shifting the neighborhood along the frequency- and phase-encoding directions, the source matrix and target matrix can be obtained, as illustrated in the right part. The parameters of the nonlinear frame can be calculated by moving the neighborhood to traverse the entire ACS region.

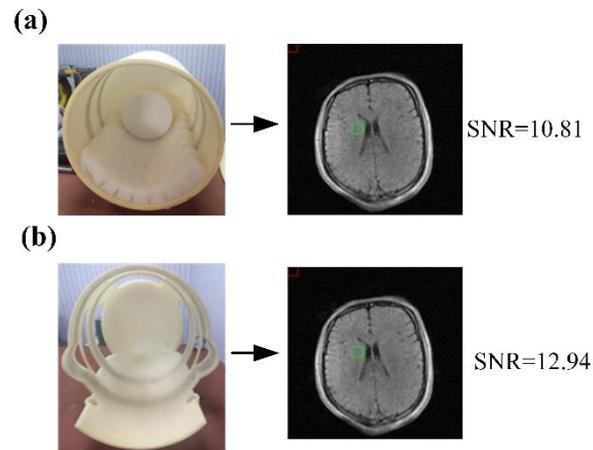

Fig. 3. The experiment of head scan for the same healthy volunteer using different RF coils, (a) the larger coil and the corresponding head scan image, the SNR of it is about 10.81, (b) the smaller coil and the head scan image of the same healthy volunteer, the SNR of it is about 12.94.

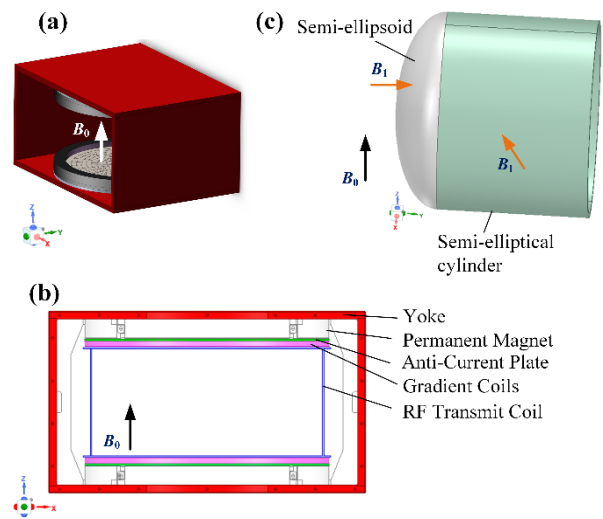

Fig. 4. Diagram of the permanent magnet of VLF MRI scanner and the skeleton of the RF coil, (a) diagram of the H-shaped permanent magnet used in our VLF MRI scanner, (b) detail information of the permanent magnet, (c) the components of the RF skeleton and the effective area.

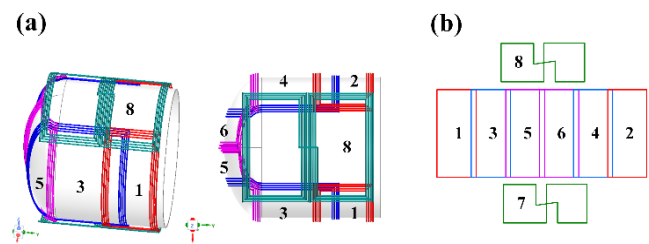

Fig. 5. Diagram of the 8-channels phased-array RF coil, (a) the 3D 8-channels phased-array RF coil and each coil unit has four turns, (b) the number and relative position of each coil unit.

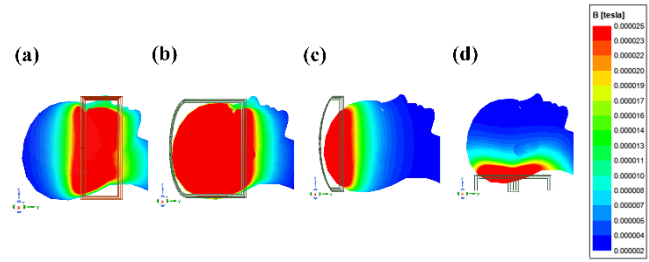

Fig. 6. The sensitivity maps of (a) CH1, (b) CH3, (c) CH5, (d) CH7.

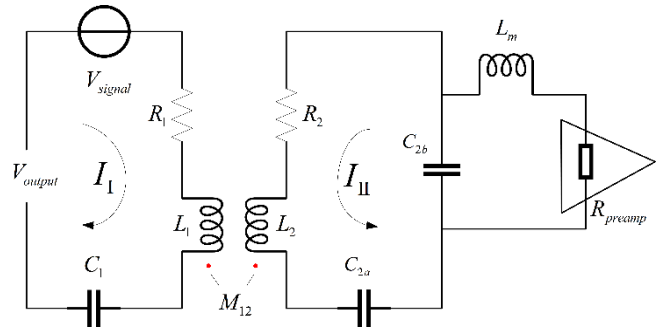

Fig. 7. Diagram of two coupled coils, and the application of LII preamplifier to prevent the crosstalk between coils.

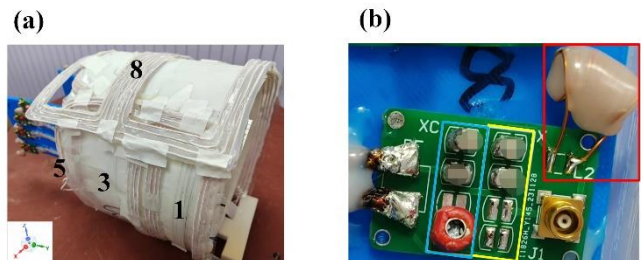

Fig. 8. (a) The prototype of 8-channels phased-array coil configuration and (b) the T&M circuit. In the T&M circuit, two non-magnetic capacitors in series are used for tuning and matching. The capacitors in the blue and yellow rectangles are $C_{2a}$ and $C_{2b}$ respectively, as shown in Fig. 7. The inductor in the red rectangle represents $L_m$ in the Fig. 7, which resonates in parallel with $C_{2b}$ at the target frequency.

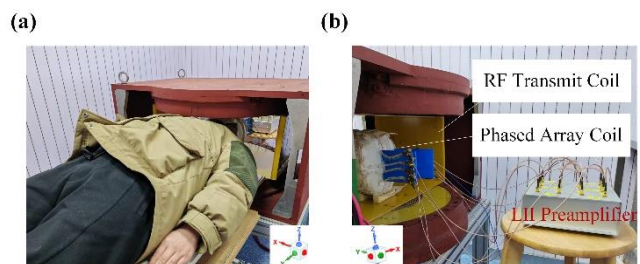

Fig. 9. The experiment platform of healthy volunteer head scanning, (a) the scanner was placed in an electromagnetic shielding room; (b) the proposed phased-array coil was placed in the RF transmit coil and connected to LII preamplifier.

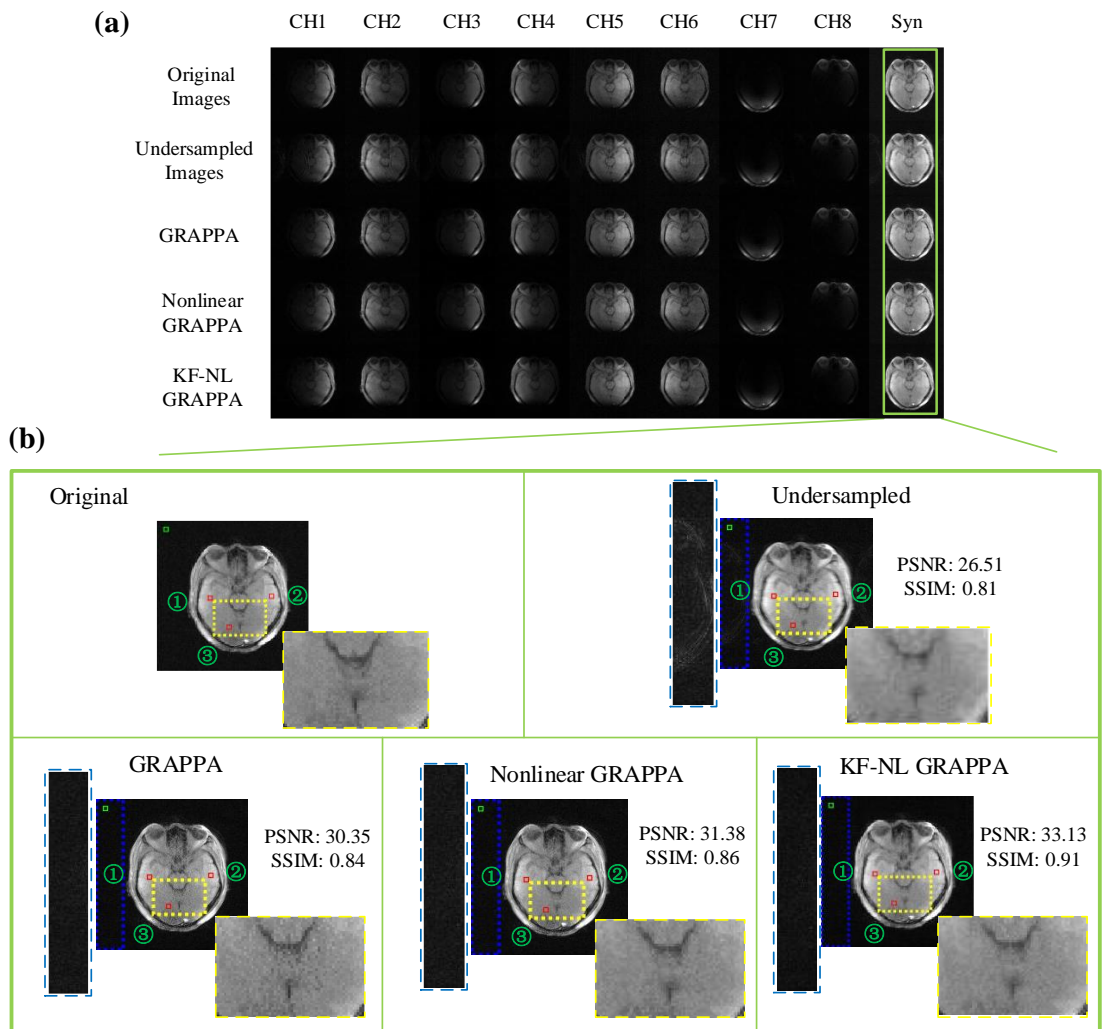

Fig. 10. The image results of healthy volunteer head scanning, (a) the image results for each coil unit, along with the synthesized images for fully sampled, undersampled, GRAPPA restored, NL-GRAPPA restored, and KF-NL-GRAPPA restored k-spaces, are presented. The term "syn" in the combined image results denotes synthesized images. Each row corresponds to different reconstruction methods, while each column represents the image results for each coil unit and the synthesized images. (b) The enlarged synthesized images are accompanied by detailed image information. The reference background region used for SNR calculation is indicated by the green square in the upper left corner, while the signal regions are marked by three red squares, each with a corresponding number in a circle. Enlarged details of the signal and background regions are provided in the yellow and blue boxes, respectively. Additionally, the peak signal-to-noise ratio and structural similarity index are also included.

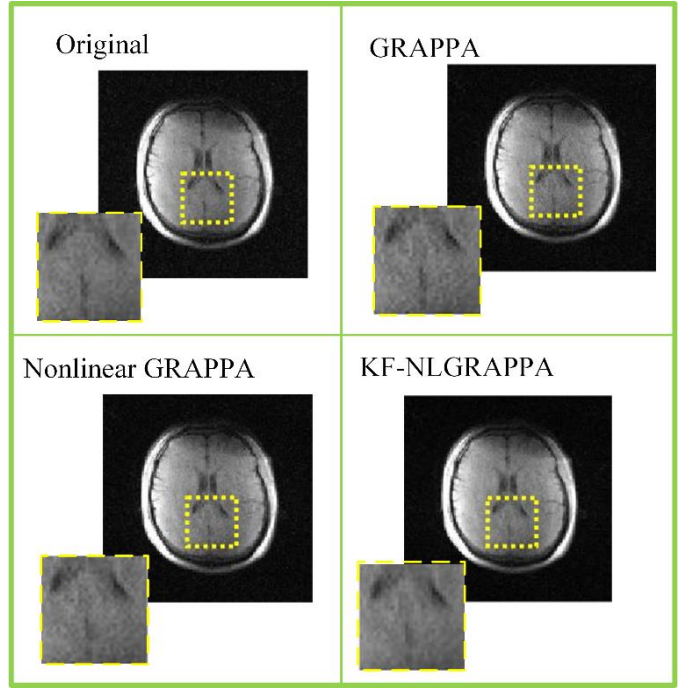

Fig. 11. The brain scan results of another healthy volunteer, the image results of fully sampled, GRAPPA restored, NL-GRAPPA restored, KF-NL-GRAPPA restored k-space.

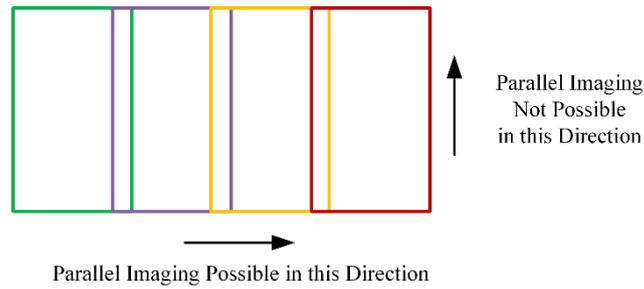

Fig. 12. Diagram of possible parallel imaging direction for a simple phased-array coil.

TABLE I
MEASURED S21 VALUES AT THE TARGET FREQUENCY FOR COIL UNITS

| S21/dB | CH1 | CH2 | CH3 | CH4 | CH5 | CH6 | CH7 | CH8 |
|---|---|---|---|---|---|---|---|---|
| **CH1** | / | -42.1 | -61.2 | -38.3 | -44.6 | -48.7 | -65.2 | -62.3 |
| **CH2** | / | / | -36.2 | -59.8 | -47.3 | -45.8 | -70.1 | -65.6 |
| **CH3** | / | / | / | -29.6 | -59.6 | -38.4 | -60.3 | -65.1 |
| **CH4** | / | / | / | / | -35.7 | -66.3 | -58.6 | -55.2 |
| **CH5** | / | / | / | / | / | -59.6 | -55.8 | -58.2 |
| **CH6** | / | / | / | / | / | / | -62.1 | -59.8 |
| **CH7** | / | / | / | / | / | / | / | -32.3 |
| **CH8** | / | / | / | / | / | / | / | / |

TABLE II
COMPARISON OF THE PERFORMANCES OF DIFFERENT RECONSTRUCTION METHODS

| Metrics | GRAPPA | NL-GRAPPA | KF-NL GRAPPA |
|---|---|---|---|
| MSE | 59.87 | 47.28 | 31.64* |
| PSNR | 30.35 | 31.38 | 33.13* |
| SSIM | 0.84 | 0.86 | 0.91* |

*, the best results

TABLE III
COMPARISON OF SNR AND TIME CONSUMPTION OF DIFFERENT IMAGE RESULTS

| Image Results | SNR1 | SNR2 | SNR3 | Time (mins) |
|---|---|---|---|---|
| Original | 10.24 | 12.72 | 6.46 | 7.20 |
| KF Original | 11.74 | 13.71 | 7.03 | 7.20 |
| Undersampled | 12.57 | 14.64 | 7.10 | 3.55* |
| GRAPPA | 11.57 | 13.63 | 6.99 | 3.55* |
| NL-GRAPPA | 11.21 | 13.06 | 6.74 | 3.55* |
| KF-NL-GRAPPA | 17.09* | 19.53* | 12.74* | 3.55* |

*, the best results